\documentclass[12pt]{article}
\usepackage[english]{babel}
\usepackage{epsfig}
\usepackage{amssymb}
\usepackage{indentfirst}
\usepackage{cite}

\renewcommand{\vec}[1]{{\bf #1}}
\def\dfrac#1#2{{\displaystyle#1\over\displaystyle#2}}

\newcommand{\ddiv}{\mathop{\rm div}\nolimits}
\newcommand{\grad}{\mathop{\rm grad}\nolimits}

\begin{document}

\begin{center}
{\bf\Large The Possible Nature of Dips in the Light Curves of Semi-Detached Binaries with
 Stationary Disks}\\[1cm]

{\it D. V. Bisikalo$^1$, P. V. Kaygorodov$^1$ , A. A. Boyarchuk$^1$, and 
O. A. Kuznetsov$^{1,2}$}\\[5mm]
{\it\small $^1$ Institute of Astronomy, Pyatnitskaya str. 48, 
 Moscow, 119017 Russia}\\
{\it\small $^2$ Keldysh Institute of Applied Mathematics, 
 Russian Academy of Sciences, Miusskaya sq. 4, Moscow, 125047 Russia}\\[1cm]
{\small\bf Published in Astronomy Reports, Vol. 49, N. 9, 2005, pp. 701-708.
Translated from Astronomicheskii Zhurnal, Vol. 82, N. 9, 2005, pp. 788-796}
\end{center}

\begin{abstract}
A thickening at the outer edge of the accretion disk is usually 
invoked to explain the dips in the light curves of cataclysmic variables with
stationary disks at phases $\sim 0.7$. The non-collisional interaction between
the stream and the disk in the stationary solution raises the question of why
matter appears at a considerable height above the accretion disk in such
systems. Our three-dimensional numerical modeling demonstrates that a
thickening of the halo above the disk can appear even in the absence of a 
direct collision between the stream and the disk. In the gas-dynamical flow
pattern described with the "hot-line" model, a considerable fraction of the
matter is accelerated in the vertical direction during the flow's interaction
with the circumdisk halo. The vertical motion of the gas due to the presence
of the $z$ component of the velocity leads to a gradual thickening of the
circumdisk halo. The computations reveal the strongest thickening of the
halo above the outer edge of the disk at phases $\sim 0.7$, in agreement with
observations for stationary-disk cataclysmic variables. This supports the
hot-line model suggested earlier as a description of the pattern of the matter
flows in semi-detached binaries, and presents new possibilities for
interpreting the light curves of such systems.
\end{abstract}

\section{Introduction}

Observations of low-mass X-ray binaries (LMXBs) have revealed
dips in the X-ray light curves for several systems. Explanations of these dips
have hypothesized the presence of a thickening at the accretion disk's outer
edge at phase $\sim 0.8$, which corresponds to the position of this feature
in the light curves ([1, 2] and references therein). The presence of matter
surrounding the X-ray source at a considerable height above the system's
orbital plane, along with the matter's uneven distribution in azimuth, can be
explained in terms of either the companion's gravitational action on the
accretion disk or the interaction between the stream of matter from the inner
Lagrangian point ($L_1$) and the disk. The coincidence of the phase of the
observed dips with the assumed position where the stream from $L_1$ approaches
the outer edge of the disk has tended to focus study on this particular region.
Beginning with [3--5], numerical studies of variations of the disk's scale
height due to its interaction with the stream were initiated. Currently, the
idea that the stream ricochets off the disk's outer edge is considered the
best way to explain the presence of matter at heights considerably exceeding
the disk thickness. The possibility that the stream flows around the edge of
the disk was first discussed in [6--9]; in 1996, the gas-dynamical computations
of Armitage and Livio [10] showed that a considerable fraction of the stream's
matter can ricochet off the edge of the accretion disk. In their model, after
colliding with the disk edge, some of the matter in the stream rises to a
considerable height (compared to the disk thickness), forming a stream
towards the inner parts of the disk. The computations of [10] demonstrated
that this stream of matter could explain the presence of the dips observed in
LMXB light curves. 

\begin{figure}
\centerline{\hbox{\epsfig{figure=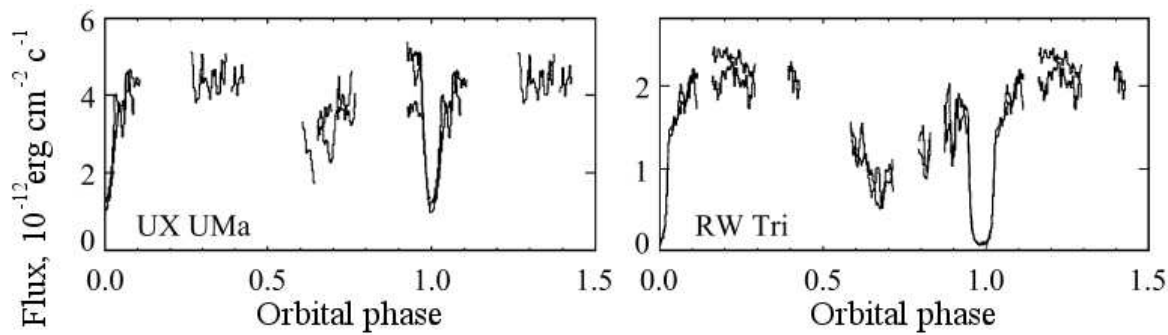,width=16cm}}}
\caption{\small Light curves of the UX UMa (left) and RW Tri (right) systems 
in the ultraviolet ($1415\div 1433 \mathrm{\AA}$),
according to [16].}
 \label{mason}
\end{figure}

It is easiest to identify dips due to the presence of
matter located high above the disk in LMXBs, since they contain a very compact
source at the center of the disk. However, similar light-curve features in
various wavelength ranges have also been recorded for a number of cataclysmic
binaries in outburst, such as U Gem [11, 12], OY Car [13, 14], and Z Cha [15].
Further observations showed that lightcurve dips can also appear when a system
is in a stationary state. Studies of the ultraviolet light curves of the
eclipsing nova-like cataclysmic binaries UX UMa and RW Tri [16] confirmed this
result, and suggested that this phenomenon was universal in semi-detached
binaries with accretion disks. As an example, Fig.~\ref{mason} presents the observed
light curves for UX UMa (left panel) and RW Tri (right panel) from [16]. It is
interesting that, in contrast to the cataclysmic systems, systems with
stationary disks display pre-eclipse dips at much earlier phases\footnote{As
usual in the analysis of observational data, the phase angle, $\phi$, is
measured from the line connecting the centers of the two stars, with the
phase increasing in the direction opposite to the system's rotation.}, 
$0.6\div 0.7$ [16, 17] (Fig.~\ref{mason}).

This raises the question of what leads to
the presence of matter at considerable heights above the accretion disk in the
case of a stationary interaction between the stream and the disk. Gasdynamical
studies of the terminal flow pattern in semi-detached binaries demonstrate that
the interaction between the stream and the disk is collisionless in this case
[18--24]. In contrast to the hot-spot model, which assumes that the stream
impacts the edge of the accretion disk, the stream interacts with gas of
the circumdisk halo in the stationary case, forming an extended region of
enhanced energy release, or so called "hot line". This means that there will
be no ricochet of the stream in the steady-state mode, and hence this model cannot
explain the dips in the light curves of binaries with stationary disks. 

Our aim here is to study possible ways of thickening the halo above a disk, which
give rise to eclipses of the central source and the presence of dips in the
light curves during the stationary flow of matter in semidetached binaries.
Our three-dimensional modeling of the flow structure demonstrates that,
during the stream's interaction with the circumdisk halo, a considerable
fraction of the matter is accelerated in the vertical direction and rises to
heights appreciably exceeding the disk scale height. The strongest thickening
of the halo above the outer edge of the disk occurs at phase $\sim 0.7$, in
agreement with observations of systems with a stationary disks.
This provides new possibilities for interpreting the light curves of such
systems. 

\section{The Model}

We used the model of [22] in our numerical study of the gas dynamics of the
matter flows in a semidetached binary. We described the flow pattern using a
three-dimensional system of gravitational gasdynamical equations including the
effects of radiative heating and cooling of the gas:

\begin{equation}
\left\{
\begin{array}{l}
\dfrac{\partial\rho}{\partial t}+\ddiv\rho\vec{v}=0\,,\\[3mm]
\dfrac{\partial\rho\vec{v}}{\partial t}
+\ddiv(\rho\vec{v}\otimes\vec{v})+\grad P=-\rho\grad\Phi
- 2 [\vec{\Omega}\times\vec{v}]\rho
\,,\\[5mm]
\dfrac{\partial\rho(\varepsilon+|\vec{v}|^2/2)}{\partial t}
+\ddiv\rho\vec{v}(\varepsilon+P/\rho+|\vec{v}|^2/2)=\\
\qquad\qquad-\rho\vec{v}\grad\Phi+\rho^2m_p^{-2}
\left(\Gamma(T,T_{wd})-\Lambda(T)\right).
\end{array}
\right. \label{HDC}
\end{equation}

Here $\rho$ is the density, $\vec{v}=(u,v,w)$ the velocity vector, $P$ the
pressure, $\varepsilon$ the internal energy, $\Phi$ the Roche potential, $m_p$
the proton mass, and $\Gamma (T,T_{wd})$ and $\Lambda(T)$ the radiative heating
and cooling functions, respectively. The system of gas-dynamical equations
was closed with the equation of state for an ideal gas,
$P=(\gamma-1)\rho\varepsilon$, where $\gamma$ is the adiabatic index. The
parameter  was taken to be $5/3$. 

We solved this system of equations using the
Roe--Osher technique [21, 25, 26] adapted for multiprocessor computers. The
modeling was carried out in a noninertial frame of reference rotating with the
binary, in Cartesian coordinates, on a rectangular three-dimensional grid. 
Since the problem is symmetric about the equatorial plane, we modeled only half
of the space occupied by the disk. The size of the modeled region, 
$1.12 A\times 1.14 A\times 0.17 A$ (where $A$ is the distance between the
system's components), was chosen to completely include the disk and the stream
of matter from the point $L_1$. The computational grid had 
$121\times 121\times 62$ cells distributed among 81 processors forming a 
$9\times 9$ two-dimensional array. The grid was made denser in the region of
interaction between the stream and the disk in order to improve the accuracy of
the solution in this region. The grid was also denser near the equatorial
plane, providing good resolution of the disk's vertical structure. 

The solution
obtained for the model without cooling was used for the initial 
conditions [27]. Before the solution converged, the model with cooling was
computed for $\approx 5$ orbital periods of the system as a whole.
The total computation
time was $\approx 1000$ hours at the MVS 1000A computer of the Joint
Supercomputer Center (JSCC). 

A free boundary condition with constant density 
$\rho_b=10^{-8}\rho_{L_1}$, a temperature of $13 600^\circ$~K, and zero
velocity was imposed at the outer boundaries of the computational domain 
(except near $L_1$ ), where $L_1$ is the matter density at $L_1$. The 
accretor was taken to be a sphere of radius $10^{-2} A$. All the matter
entering the accretor cells was taken to fall onto the star. The stream was
specified as a boundary condition: matter with temperature $5800^\circ$~K,
density $\rho_{L_1}=1.6\times 10^{-8} g/cm^3$ , and velocity along the 
$x$ axis
$v_x = 6.3 km/s$ was injected into a region with radius $0.014 A$ around $L_1$.

We considered a semi-detached binary consisting of a donor with mass $M_2$
that fills its Roche lobe and an accretor with mass $M_1$. The following
parameters were adopted for the system: $M_1=1.02 M_\odot, M_2=0.5 M_\odot$
, and $A=1.42 R_\odot$ , corresponding to $P_{orb} = 3.79^h$. The disk in the
model had a temperature of $13 600^\circ K$. For the specified rate of matter
entering the system, the corresponding accretion rate in the model was 
$\approx 10^{-10} M_\odot/yr$. 

\section{Results of the Computations}
\label{results}

\begin{figure}
\centerline{\hbox{\epsfig{figure=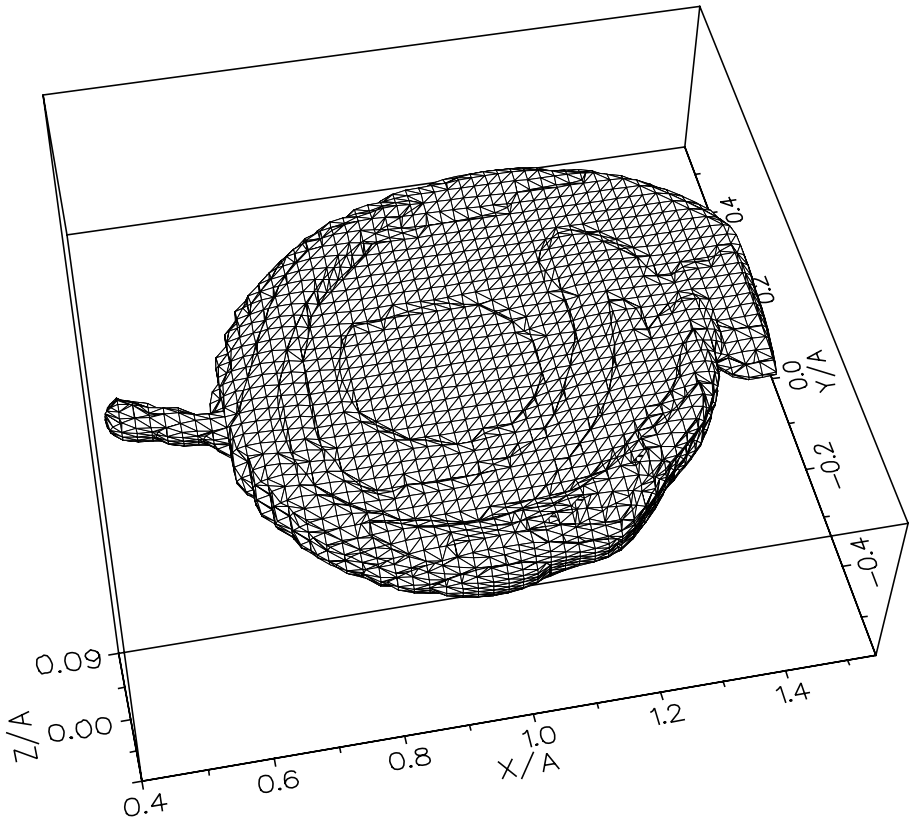,width=12cm}}}
\caption{\small Three-dimensional surface of constant density 
$\rho=5\times10^{-11} g/cm^3$.}
\label{fig_3drho}
\end{figure}

\begin{figure}
\begin{center}
\begin{tabular}{cc}
\epsfig{figure=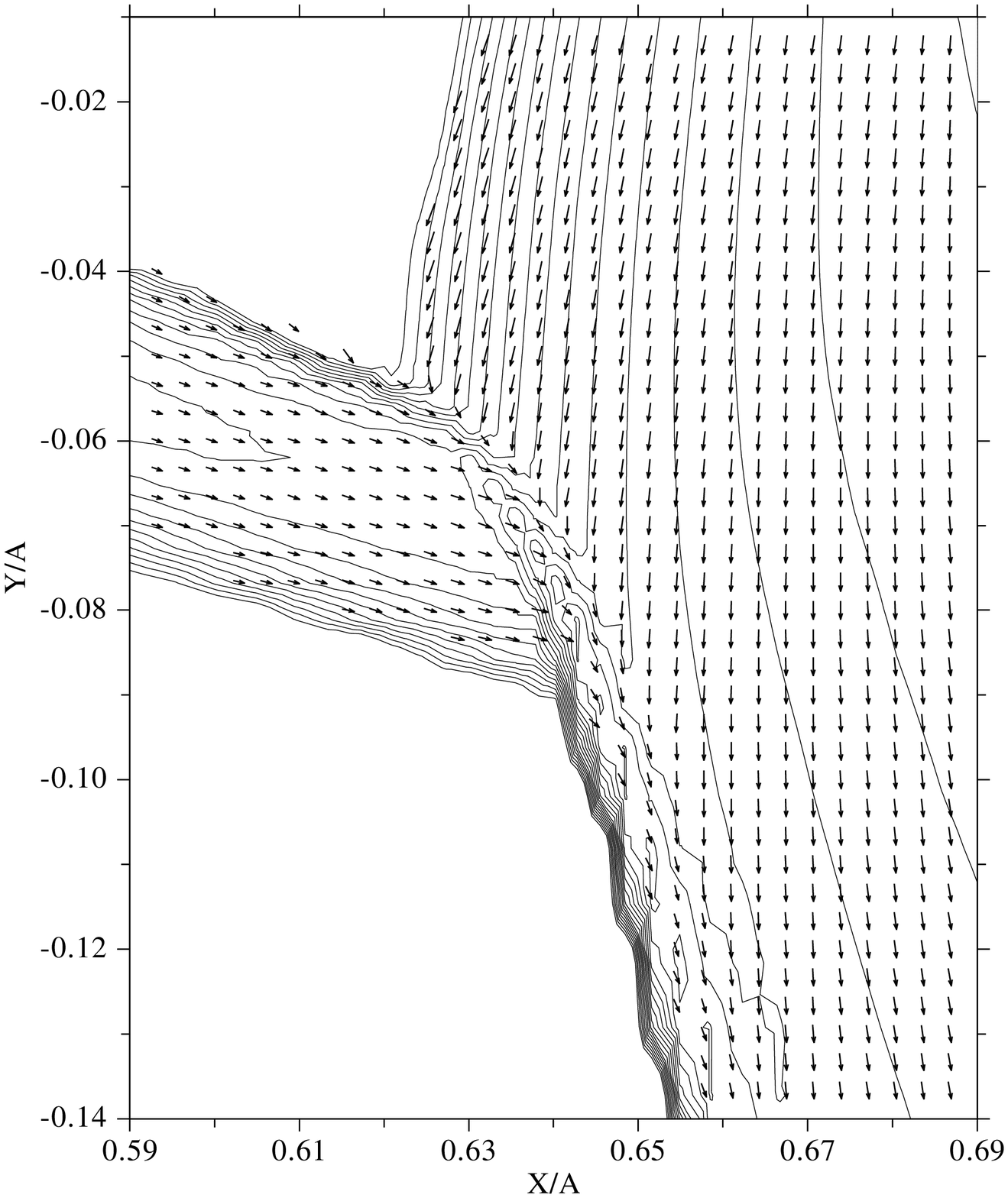,height=7cm}&
\epsfig{figure=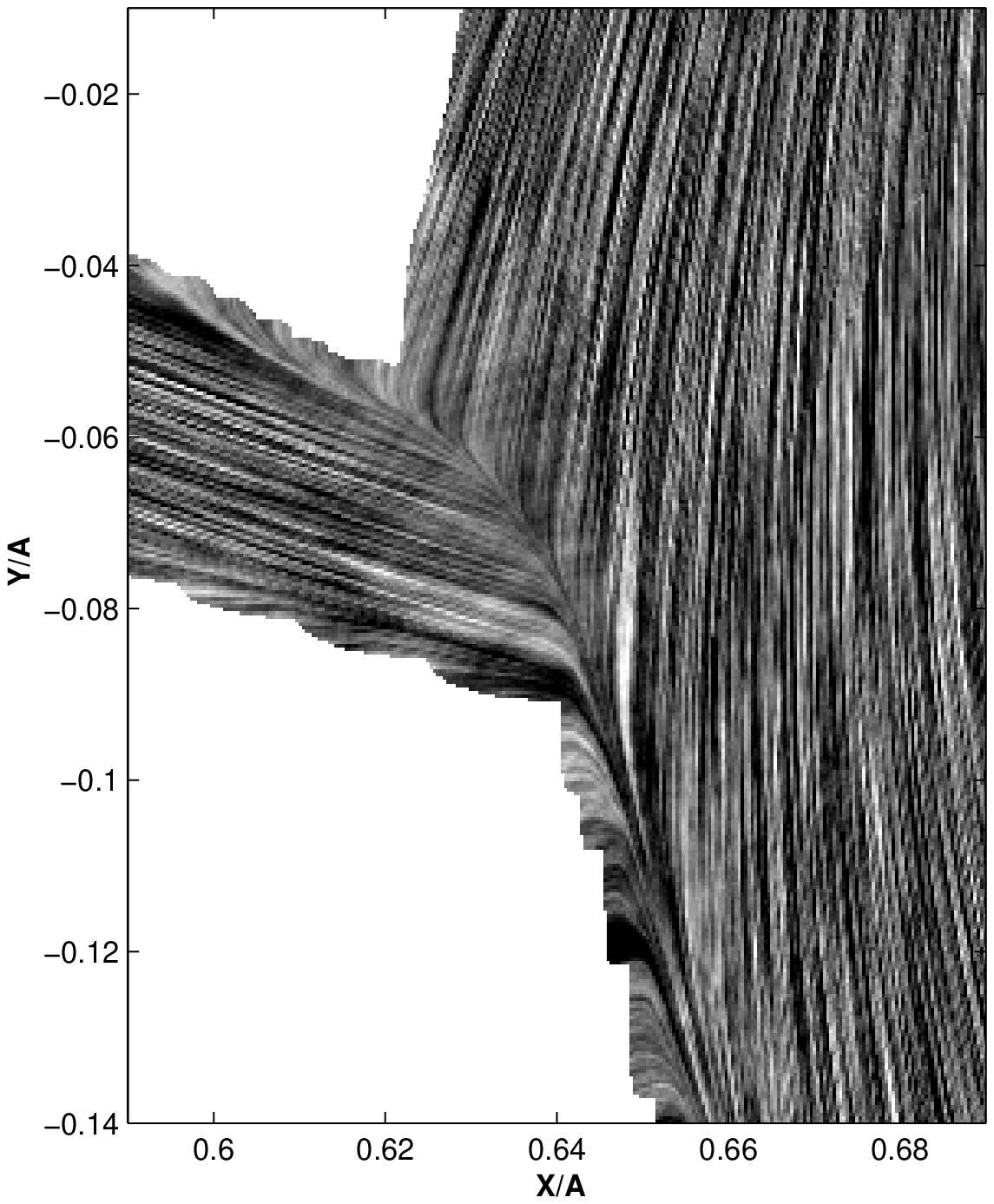,height=7cm}
\end{tabular}
\end{center}
\caption{\small Contours of constant density and velocity vectors (left panel) and a
visualization of the velocity field (right panel) in the region of
interaction between the stream and the circumdisk halo, in the system's
equatorial plane. This figure was first published in [22].}
\label{fig78}
\end{figure}

We described the morphology of the matter flows in a semi-detached binary with
a stationary, cool ($T=13 600^\circ K$) disk in [22]. Let us briefly 
summarize the main features of the computed flow structure.
Figure~\ref{fig_3drho} shows 
the three-dimensional surface of constant density 
$\rho=5\times 10^{-11} g/cm^3$.
The region of the interaction between the stream and circumdisk halo is shown
enlarged in Fig.~\ref{fig78} [22, Figs. 7, 8]. The left panel of 
Fig.~\ref{fig78} displays 
contours of constant density and velocity vectors, while the right panel of
Fig.~\ref{fig78} is the so-called texture: a visualization of the velocity
field by means
of numerous tracks of test particles. 

The results presented demonstrate that 
the interaction between the circumdisk halo and the stream possesses all the
characteristic features of an oblique collision of two flows. The resulting 
structure of two shock waves with a tangential discontinuity between them is 
clearly visible in Fig.~\ref{fig78}. The region of the shock interaction
between the 
stream and halo has a complex shape. The parts of the halo far from the
disk have low density, and the shock due to their interaction with the stream
lies along the edge of the stream. The shock bends as the gas density in the
halo increases, finally assuming a position along the edge of the disk. Note
that the solution for the cool case has the same qualitative features as the
solution for hot outer parts of the disk [27]: the interaction between the 
stream and the disk is collisionless, the region of enhanced energy release 
is due to the interaction between the gas in the circumdisk halo and in the 
stream and is located outside the disk, and the resulting system of shocks is 
extended and can be considered as a "hot line". 

It follows from the above general
features of the flow pattern that, at the interaction zone, the halo gas and
stream gas pass through the shocks corresponding to their own flows, are mixed,
and then move along the tangential discontinuity between the two shocks. 
Subsequently, the disk itself, halo, and inter-component envelope are formed
of precisely this matter. The jump in the gas parameters after the passage of
the shock leads to a the increase of density and temperature in the region
between the shocks and,
consequently, to the appearance of a pressure gradient along the $z$ axis,
perpendicular to the system's plane of rotation. As a result, the gas begins
to expand vertically, increasing the $z$ component of the velocity, until the 
pressure gradient is balanced by the gravitational force.

\begin{figure}
\begin{center}
\epsfig{figure=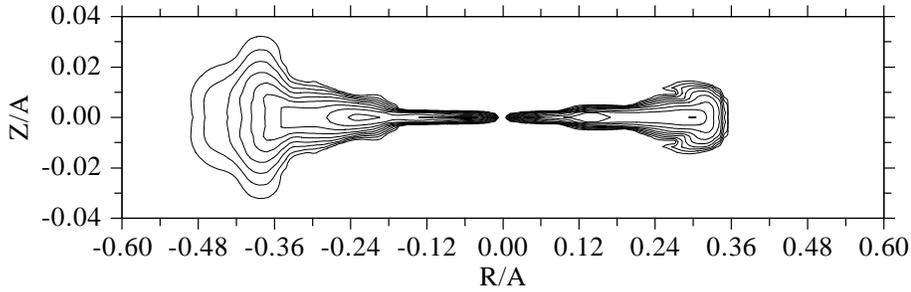,width=12cm}
\end{center}
\caption{\small Contours of constant density in the section of the flow structure
perpendicular to the equatorial plane and passing through the accretor and
the region of the halo's maximum thickening ($\phi = 0.7$).
}
\label{fig6}
\end{figure}

\begin{figure}
\centerline{\hbox{\epsfig{figure=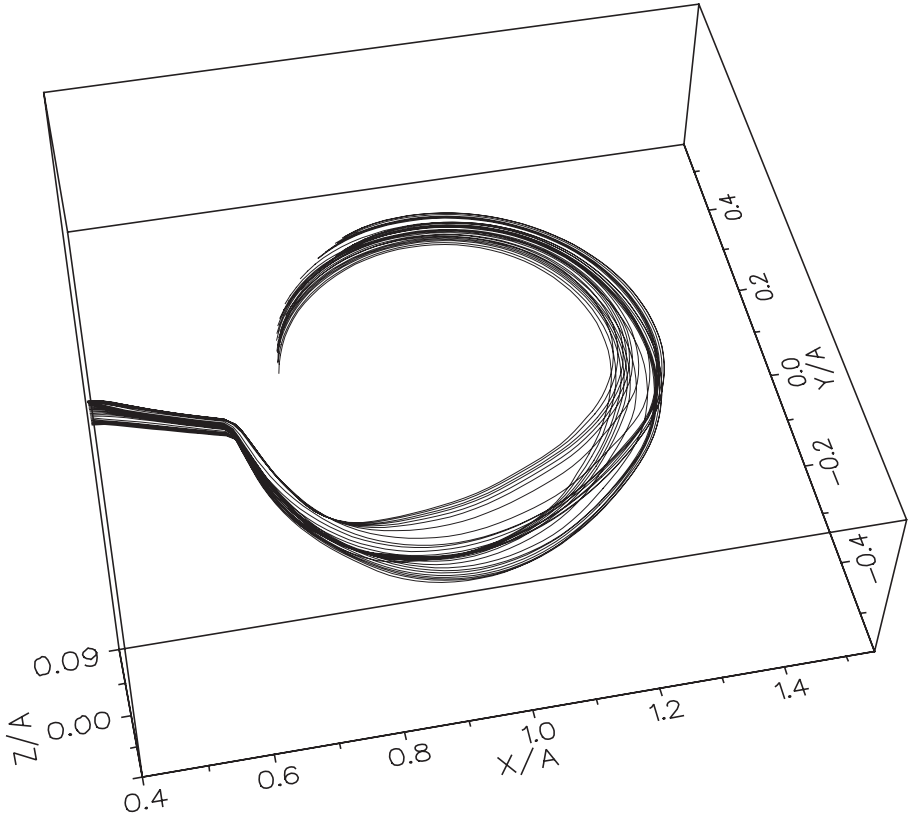,width=12cm}}}
\vspace{5mm}
\caption{\small Fragments of streamlines emerging from the neighborhood of $L_1$.}
\label{fig_tracs}
\end{figure}

The vertical gas 
pressure due to the presence of the $z$ component of the velocity, together 
with the motion of the gas along the tangential discontinuity at the disk's 
outer edge, lead to a gradual increase of the thickening of the circumdisk 
halo (along the $z$ axis). This thickening of the halo along the outer edge of
the accretion disk is clearly visible in the three-dimensional 
constant-density surface shown in Fig.~\ref{fig_3drho}, as well as in 
Fig.~\ref{fig6}, which displays
constantdensity curves in the section of the flow structure that is 
perpendicular to the equatorial plane and drawn through the accretor and the
maximum thickening of the halo ($\phi=0.7$). An additional illustration of
the halo thickening above the disk is provided by Fig.~\ref{fig_tracs}, which
shows fragments
of computed streamlines emerging from the vicinity of $L_1$. We can see that
the streamlines diverge after acquiring a vertical acceleration in the region
between the shocks, then converge again toward the system's equatorial plane.
The region of vertical acceleration is restricted to the hot-line zone. After
passing this region, the gas has a large vertical velocity component that 
makes it climb higher, until its store of kinetic energy is exhausted. The
point where the upward motion ceases corresponds to the maximum height, which
is reached at phase $\sim 0.7$, i.e., already substantially outside the hot
line. 

\begin{figure}
\centerline{\hbox{\epsfig{figure=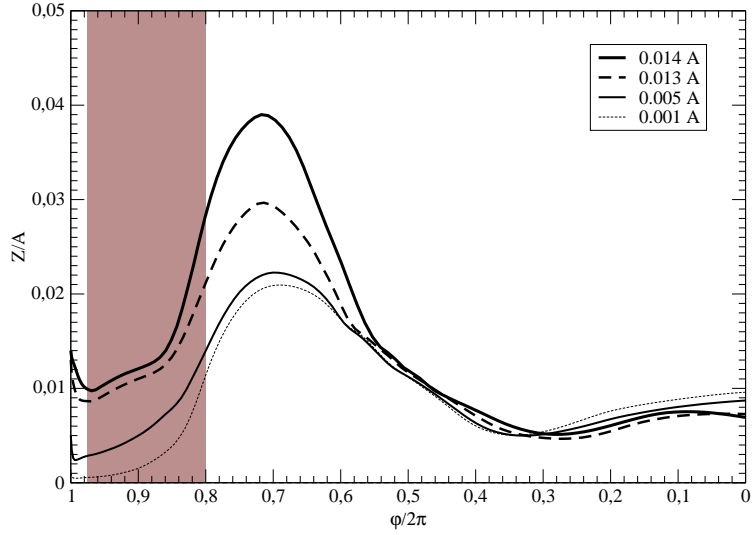,height=7cm}}}
\vspace{5mm}
\caption{\small $z(\phi)$ relation for four streamlines. The streamlines originate in
the vicinity of $L_1$ and have the coordinates $(x_{L_1} ,0, z_0)$. The $z_0$
values for each streamline are indicated in the plot. The shaded area
corresponds to the region where $v_z$ increases.}
\label{fig_z}
\end{figure}

\begin{figure}
\centerline{\hbox{\epsfig{figure=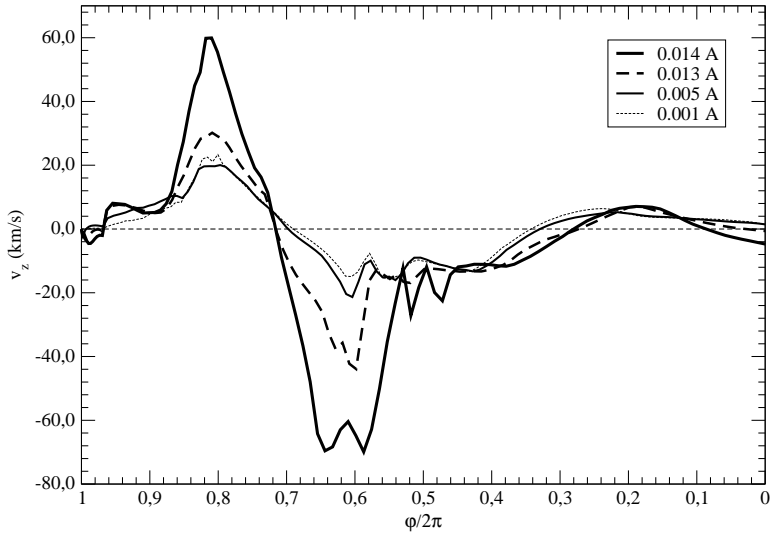,height=7cm}}}
\vspace{5mm}
\caption{\small $V_z(\phi)$ relation along the same four streamlines as in
Fig.~\ref{fig_z}.}
\label{fig_vz}
\end{figure}

To quantitatively analyze the thickening of the halo above the disk,
let us consider the behavior of the streamlines and the distributions of the
gas parameters along them. In a cylindrical coordinate system with its origin
coincident with the accretor $(x=A, y=0.0, z=0.0)$ and the angle $\phi$ 
measured from the point $L_1$ opposite to the direction of rotation of the
matter (coincident with the system's direction of rotation), each point of
the streamline is described by the coordinates $(r, \phi, z)$. The $z(\phi)$
relations for four streamlines originating at points in the neighborhood of 
$L_1$ are presented in Fig.~\ref{fig_z}. We can see that, after entering the
hot-line
region (phase $\sim 0.975$), the streamlines begin to climb due to the
increase of the velocity's $z$ component. The phase relation of the vertical
velocity for the same four streamlines is displayed in Fig.~\ref{fig_vz}.
When the gas
emerges from the hot line (phase $\sim 0.8$), the force due to the pressure
jump behind the shocks disappears, and the gas simply moves in the accretor's
gravitational field. The vertical velocity of the gas becomes zero at phase 
$\sim 0.7$, corresponding to the maximum ascent of the streamlines. The height
of the halo at this position reaches $\sim 0.04 A$.

In addition to the
primary maximum, the system also exhibits a minimum height at phase $\sim 0.3$,
when $z\sim(0.005\div0.006)A$, and a secondary maximum at phase $\sim 0.1$,
when the height of the halo is $\sim 0.01 A$. Due to the effects of viscous 
dissipation, the minimum and secondary maximum are considerably less pronounced
than the primary maximum. Eventually, viscosity leads to complete damping of
the vertical velocity oscillations, so that the gas no longer possesses a 
significant vertical velocity when it next approaches the region of the 
interaction with the stream.

\begin{figure}
\centerline{\hbox{\epsfig{figure=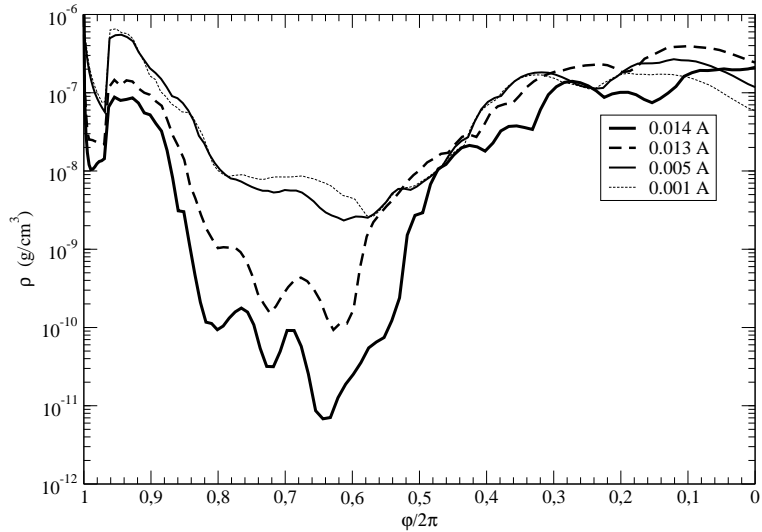,height=7cm}}}
\vspace{5mm}
\caption{\small The density relation $\rho(\phi)$ along the same four
streamlines as in Fig.~\ref{fig_z}.}
\label{fig_rho}
\end{figure}

Figure~\ref{fig_rho} is the phase dependence of the density 
for the same four streamlines as in Fig.~\ref{fig_z}. The densities
correspond to the 
mass-exchange rate in the RW Tri system ($10^{-8} M_{\odot}/yr$ [28]).
We can see
from these results that a strong density decrease is observed in the region of
maximum ascent of the streamlines. This behavior of the density can lead to a
displacement of the light-curve dips towards regions of higher $\rho$; i.e.,
towards
higher phases. This should be manifest most strongly in short-wavelength 
observations, since higher densities are required to absorb harder radiation. 
It is also interesting that the gas density is much higher near the secondary 
maximum, at phases $\sim 0.1$, than near the first thickening.

\begin{figure}
\centerline{\hbox{\epsfig{figure=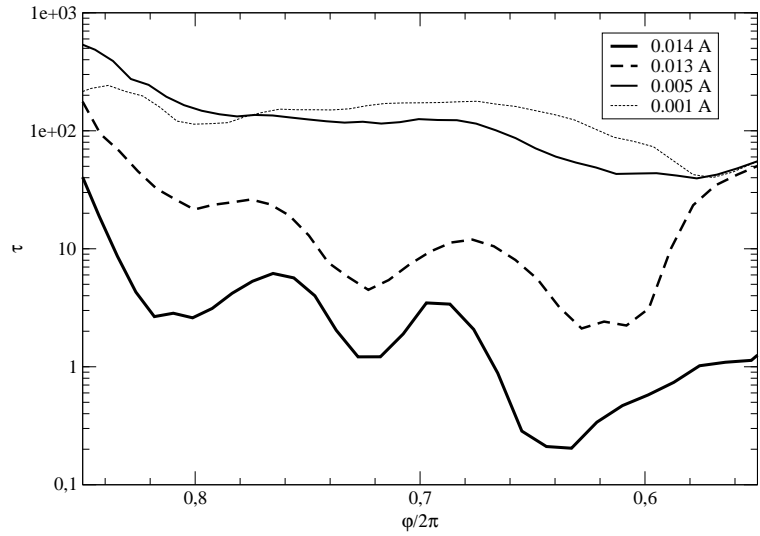,height=7cm}}}
\vspace{5mm} \caption{\small The optical depth relation, $\tau(\phi)$,
near the thickening ($\phi=0.55\div 0.85$) along the same four streamlines
as in Fig.~\ref{fig_z}.}
\label{fig_tau}
\end{figure}

To evaluate 
whether the central source can be eclipsed by the thickening of the halo, we 
computed the optical depth of the halo matter, $\tau$. The optical depth is 
the product of the density, the layer's geometrical depth, and the absorption 
coefficient: $\tau=\rho l\kappa$. To estimate $\tau$ near the region of
thickening, let us use
the densities presented in Fig.~\ref{fig_rho}, taking the characteristic linear
size of the
thickening to be its halfheight (Fig.~\ref{fig_z}). We estimate the absorption
coefficient using the approximation formulas from [29] with a temperature of 
$\sim 13 600^\circ K$ and a characteristic density corresponding to the value
at the
thickening's half maximum. Our estimates of $\tau$ are presented in
Fig.~\ref{fig_tau}, which shows
the relations between the optical depth and the phase, $\tau(\phi)$, along the
same four streamlines as in Fig.~\ref{fig_z}. An analysis of these estimates
of $\tau$ 
shows that the thickening is optically thick, even at its greatest heights,
and the value of $\tau$ at the half maximum can reach $\sim 10^2$.
 This means that
the halo thickening is capable of eclipsing the central source, giving rise to
the dips observed in the light curves of semi-detached binaries with stationary
disks. 

\begin{figure}
\centerline{\hbox{\epsfig{figure=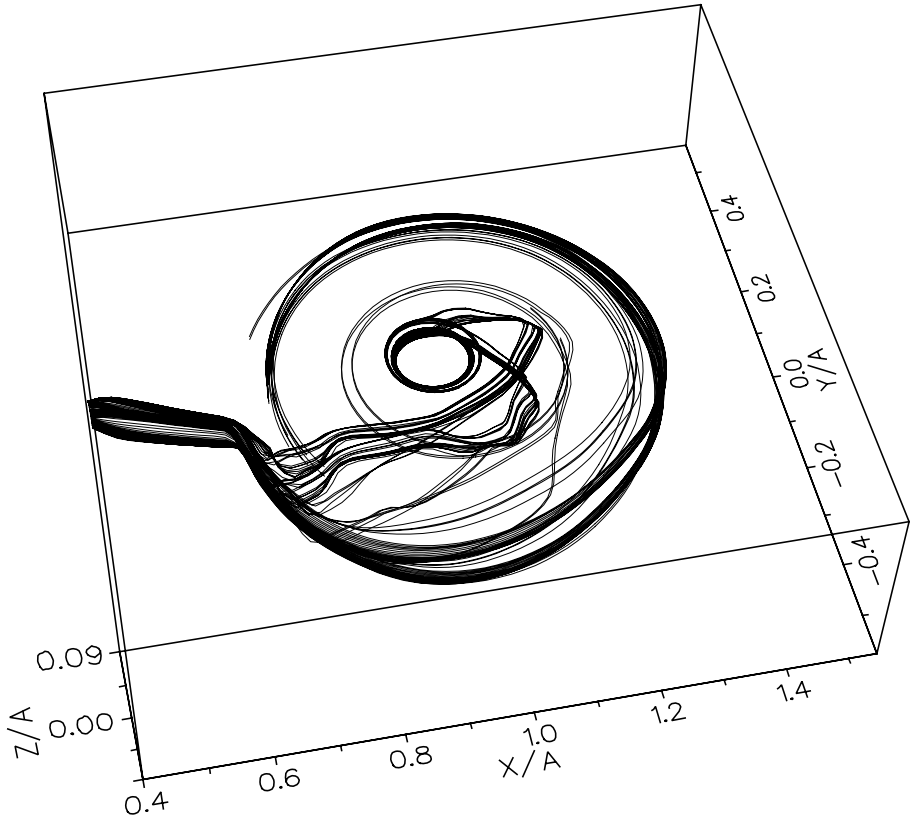,width=12cm}}}
\vspace{5mm}
\caption{\small 
Same as Fig.~\ref{fig_tracs}, also showing fragments of the streamlines
emerging from a wider vicinity of $L_1$.}
\label{fig_tracs2}
\end{figure}

Obviously, in stationary systems, when there is no collision between the
stream and the outer edge of the disk, there can be no ricochet and no transfer
of a substantial fraction of the stream's matter into the inner parts of the
disk. However, the momentum of the stream matter is lower than that of the halo
gas, so that some of the stream gas at great heights can slide down into inner
orbits, appearing as if the stream is flowing around the outer edge of the
disk. This sliding-down can take place only in the hotline region, where the
momenta of the matter in the stream and in the halo have not yet become
equalized. Indeed, when considering the streamlines that reach high heights
from the stream's center (Fig.\ref{fig_tracs2}),
we can see that some of them leave the
interaction region of the stream and halo, envelop the disk from the upper
side, and get into its inner regions. The results of our computations
demonstrate that the total flux of matter associated with this effect is small,
and apparently cannot significantly influence the visibility of the central
object. 

\section{Conclusions}

Observations of semi-detached binaries with stationary accretion disks reveal
dips in their light curves at phase $\sim 0.7$. These dips are usually
explained by invoking the presence of a thickening at the outer edge of the
accretion disk, without explaining how matter appears at considerable heights
above the accretion disk in the case of a stationary interaction between the
stream and disk. It is important to resolve the problem of the formation of
this thickening of the halo above a stationary disk if we wish to correctly
interpret the observations and more fully understand the nature of 
matter-exchange processes in these stars. 

The results of our three-dimensional
numerical modeling of the matter-flow structure in semidetached binaries with
stationary disks confirm our earlier conclusions [21, 22, 27] that the
interaction between the stream and disk is collisionless, the region of
increased energy release is due to the interaction between the gas in the
circumdisk halo and the stream and is located outside the disk, and the system
of shocks that forms is extended and can be considered as a "hot line." The 
interaction between the circumdisk halo and the stream possesses all
the characteristic features of an oblique collision of two flows that results
in the formation of a structure consisting of two shock waves and a tangential
discontinuity.

In the hot-line region, the halo and stream gases pass through 
the shocks corresponding to their own flows, are mixed, and then move along 
the tangential discontinuity between the two shocks. During the interaction 
between the stream and the circumdisk halo, a considerable fraction of the 
matter acquires a vertical acceleration. The vertical motion of the gas due 
to the $z$ component of the velocity, together with its motion along the 
tangential discontinuity at the outer edge of the disk, results in a gradual 
growth of the thickness of the circumdisk halo. The region of vertical 
acceleration is restricted to the hot-line zone, and its angular size does not
exceed $\sim 65^\circ$.
However, once it has passed this region, the gas has a 
sufficiently large vertical velocity component to rise until its store of 
kinetic energy is exhausted. The point where the upward motion ceases 
corresponds to the maximum height, which is reached at phase $\sim 0.7$,
already substantially outside the hot line. The thickening extends appreciably
higher than the scale height of the disk, reaching values $\sim 0.04 A$ (this
corresponds to a ratio of the thickening's height to the distance to the 
accretor $>0.1$), and its angular size exceeds $\sim 130^\circ$.
Our computations
also show that, in addition to the primary maximum, the system also displays
a height minimum at phase $\sim 0.3$, when $z\sim (0.005\div0.006)A$, and a
secondary maximum at phase $\sim 0.1$, when the height of the halo is 
$\sim 0.01 A$.

Our analysis of these results leads us to conclude that the
dips in the light curves of semi-detached binaries with stationary disks 
(i.e., in the absence of a collisional interaction between the stream and
disk), can be explained by the formation of a thickening of the halo above
the outer edge of the disk. The origin of this thickening is described well
by the hot-line model, and its quantitative features are consistent with 
observations. The proposed model can be applied to both semi-detached systems
(LMXBs, cataclysmic variables) in their stationary state and to dwarf novae
in outburst, provided that the outer parts of the disk are not strongly 
distorted in the outburst. 

\section{Acknowledgments}

This study was supported by the Russian Foundation for Basic Research (project
nos. 03-0216622, 03-01-00311, 05-02-16123, 05-02-17070, and 05-02-17874), the
Program of Support for Leading Scientific Schools of Russia (grant no.
NSh162.2003.2), and the basic-research programs "Mathematical Modeling and
Intellectual Systems" and "Unstable Phenomena in Astronomy" of the Presidium
of the Russian Academy of Sciences. 

\section{References}

1. N. E. White and S. S. Holt, Astrophys. J. 257, 318 (1982). 

2. K. O. Mason, in Two Topics in X-ray Astronomy, Vol. 1: X-ray Binaries, ESA SP-296, p. 113 (1989). 

3. M. Livio, N. Soker, and R. Dgani, Astrophys. J. 305, 267 (1986). 

4. R. Dgani, M. Livio, and N. Soker, Astrophys. J. 336, 350 (1989). 

5. M. Hirose, Y. Osaki, and S. Mineshige, Publ. Astron. Soc. Jpn. 43, 809 (1991). 

6. S. H. Lubow and F. H. Shu, Astrophys. J. 198, 383 (1975). 

7. S. H. Lubow and F. H. Shu, Astrophys. J. 207, L53 (1976). 

8. S. H. Lubow, Astrophys. J. 340, 1064 (1989). 

9. J. Frank, A. R. King, and J.-P. Lasota, Astron. Astrophys. 178, 137 (1987).

10. P. J. Armitage and M. Livio, Astrophys. J. 470, 1024 (1996). 

11. K. O. Mason, F. A. Cordova, M. G. Watson, and A. R. King, Mon. Not. R. Astron. Soc. 232, 779 (1988).

12. K. S. Long, C. W. Mauche, J. C. Raymond, et al., Astrophys. J. 469, 841 (1996). 

13. T. Naylor, G. T. Bath, P. A. Charles, et al., Mon. Not. R. Astron. Soc. 231, 237 (1988). 

14. I. Billington, T. R. Marsh, K. Horne, et al., Mon. Not. R. Astron. Soc. 279, 1274 (1996). 

15. E. T. Harlaftis, B. J. M. Hassall, T. Naylor, et al., Mon. Not. R. Astron. Soc. 257, 607 (1992). 

16. K. O. Mason, J. E. Drew, and C. Knigge, Mon. Not. R. Astron. Soc. 290, L23 (1997). 

17. C. S. Froning, K. S. Long, and C. Knigge, Astrophys. J. 584, 433 (2003). 

18. D. V. Bisikalo, A. A. Boyarchuk, O. A. Kuznetsov, and V. M. Chechetkin, Astron. Zh. 74, 880 (1997) [Astron. Rep. 41, 786 (1997)]. 

19. D. V. Bisikalo, A. A. Boyarchuk, O. A. Kuznetsov, et al., Astron. Zh. 75, 40 (1998) [Astron. Rep. 42, 33 (1998)]. 

20. D. V. Bisikalo, A. A. Boyarchuk, V. M. Chechetkin, et al., Mon. Not. R. Astron. Soc. 300, 39 (1998). 

21. A. A. Boyarchuk, D. V. Bisikalo, O. A. Kuznetsov, and V. M. Chechetkin, Mass Transfer in Close Binary Stars (Taylor and Frances, London, 2002). 

22. D. V. Bisikalo, A. A. Boyarchuk, P. V. Kaigorodov, and O. A. Kuznetsov, Astron. Zh. 80, 879 (2003) [Astron. Rep. 47, 809 (2003)]. 

23. M. Makita, K. Miyawaki, and T. Matsuda, Mon. Not. R. Astron. Soc. 316, 906 (2000). 

24. K. Sawada and T. Matsuda, Mon. Not. R. Astron. Soc. 255, 17P (1992). 

25. P. L. Roe, Ann. Rev. Fluid Mech. 18, 337 (1986). 

26. S. R. Chakravarthy and S. Osher, in Proceedings of the 23rd Aerospace Sci. Meeting, AIAA-85-0363, p. 363 (1985). 

27. O. A. Kuznetsov, D. V. Bisikalo, A. A. Boyarchuk et al, Astron. Zh. 78, 997 (2001) [Astron. Rep. 45, 872 (2001)]. 

28. P. J. Groot, R. G. M. Rutten, and J. van Paradijs, Astron. Astrophys. 417, 283 (2004). 

29. K. R. Bell and D. N. C. Lin, Astrophys. J. 427, 987 (1994).

\end{document}